\documentclass[aps,twocolumn,superscriptaddress,groupedaddress,nofootinbib]{revtex4}  
\usepackage{graphicx}  
\usepackage{dcolumn}   
\usepackage{bm}        
\usepackage{amssymb}   
\usepackage{amsmath}   
\usepackage{multirow}
\usepackage{hyperref}
\usepackage{braket}
\usepackage{subfigure}
\usepackage{array}
\usepackage[math]{cellspace}
\usepackage[T1]{fontenc}
\usepackage{eepic,epic}
\usepackage[normalem]{ulem}

\usepackage[T1]{fontenc}
\usepackage{bbm}
\usepackage{picture}

\usepackage{color}
\usepackage[normalem]{ulem}

\def\tr#1{{\mathrm{tr}\left[#1\right]}}

\def\tab#1{Table~\ref{#1}}

\usepackage[utf8]{inputenc} 

\begin{document}

\title{Singlet channel scattering in a Composite Higgs model on the lattice}
{\flushleft{CERN-TH-2021-221}}

\date{\today}

\author{Vincent Drach}

\affiliation{Centre for Mathematical Sciences, University of Plymouth, Plymouth, PL4 8AA, United Kingdom}

\author{Patrick Fritzsch}
\affiliation{Centre for Mathematical Sciences, University of Plymouth, Plymouth, PL4 8AA, United Kingdom}

\affiliation{School of Mathematics, Trinity College Dublin, Dublin 2, Ireland}

\author{Antonio Rago}

\affiliation{Centre for Mathematical Sciences, University of Plymouth, Plymouth, PL4 8AA, United Kingdom}
\affiliation{CERN, Theoretical Physics Department, 1211 Geneva 23, Switzerland}

\author{Fernando Romero-L\'opez}
\affiliation{IFIC (CSIC-UVEG), 46980 Paterna, Spain}


\begin{abstract}
We present the first calculation of the scattering amplitude in the
singlet channel beyond QCD. The calculation is performed in $SU(2)$
gauge theory with $N_f=2$ fundamental Dirac fermions and based on a
finite-volume scattering formalism. The theory exhibits a $SU(4) \to
Sp(4)$ chiral symmetry breaking pattern that is used to design
minimal composite Higgs models currently tested at the LHC.  Our
results show that, for the range of underlying fermion mass considered, the lowest
flavour singlet state is stable. 
\end{abstract}
\maketitle

\section{Introduction}
\label{sec:intro}

The discovery of the Standard Model's (SM) last missing piece, the Higgs
boson, and the increase in precision of tests of its properties, continue to
trigger the study of numerous mechanisms to address the
fundamental problems with its formulation.

Among  other possibilities,  a new strongly interacting sector giving
rise to the observed phenomenology at the electroweak scale (EW) and below
has been pursued for decades. Such a new sector could feature a solution to the naturalness problem and provide a mechanism
to generate a non-trivial mass spectrum together with a large scale separation. These
mechanisms have been used  for instance in  the context of 
Composite Higgs
models~\cite{Terazawa:1976xx,Terazawa:1979pj,Kaplan:1983fs,Kaplan:1983sm,Dugan:1984hq,Bardeen:1985sm,Leung:1985sn,Yamawaki:1985zg}, of scenarios of dynamical electroweak symmetry
breaking~\cite{Weinberg:1975gm,Susskind:1978ms}, and of Dark Matter
models~\cite{Hochberg:2014dra,Tsai:2020vpi}. These appealing ideas motivate the lattice endeavour to understand gauge theories beyond QCD.

One feature of a strongly interacting sector is the inevitable presence
of a flavour singlet state of positive parity---referred to
as $\sigma$  in the rest of this paper. In QCD-like theories, the
$\sigma$  is expected to be a resonance of two Goldstone
bosons in the limit of massless
underlying fermions. 

{In Composite Higgs scenarios, the embedding of the new strong
  sector in the Standard Model is such that the  Goldstone bosons of the strong sector play the role
of the SM Higgs field.  
In these models, aside from the Goldstone bosons,  also the
presence of new resonances like the $\sigma$ can affect
the predictions for the LHC \cite{Contino:2011np}, and could be detected by the next generation of
colliders \cite{Bharucha:2020bhy}.
In general, the  phenomenological implications of the new scalar resonance in a
  composite Higgs scenario will depend on the underlying dynamics and on
  the details of the electroweak embedding. Unless  the model features a parametrically large scale
  separation between the Goldstone bosons and the $\sigma$,  the effective description at the
  EW scale must take the $\sigma$ into account.
The mixing of the scalar $\sigma$ resonance with the Goldstone
  bosons associated with the spontaneous symmetry breaking of the new
  strong sector is induced by the interaction with the SM, and gives rise to an additional effective scalar field
  with a larger mass, see Refs.~\cite{Bellazzini:2015nxw,Bizot:2016zyu,Niehoff:2016zso}.
  Such a resonance is expected to be produced at the LHC similarly
to the SM Higgs, i.e. via gluon fusion and vector boson fusion
  mechanisms, as discussed for instance in
  Ref.~\cite{BuarqueFranzosi:2018eaj}.
  Effective theories at the EW scale will encode the actual
  realisation of the Composite Higgs scenarios
  through their low energy constants, under the assumption of a strong
  sector weakly coupled to the Standard Model.
  Given the large number of low energy couplings parametrising the effective Lagrangian, additional theoretical constraints are needed to discriminate among Composite Higgs models. Lattice calculations
  can reduce the space of parameters by performing measurements on
  the new strong sector in isolation. The present work contributes to 
  our understanding of the role of the $\sigma$ resonance in the 
  phenomenology of the class of composite models characterised by the strong sector we are considering, irrespectively of its embedding.}

In lattice simulations the only rigorous approach to reveal the nature
of a resonance is to estimate the scattering amplitude of the
Goldstone bosons. Lattice simulations in various gauge theories have estimated the mass
of the $\sigma$ in a regime where it is stable~\cite{LatKMI:2013bhp,LatKMI:2014xoh,Fodor:2015vwa,Brower:2015owo,Arthur:2016ozw,Hasenfratz:2016gut,Appelquist:2016viq,Athenodorou:2017dbf,LatticeStrongDynamics:2018hun,Lee:2018ztv,Bennett:2019jzz}. Scattering
amplitudes have been evaluated also for other channels, see for instance the recent
work in a possible nearly conformal theory for $SU(3)$ with $N_f=8$ flavours in the maximal-isospin
channel \cite{LSD:2021xlp}  and  our recent work in the vector meson channel for
$SU(2)$ with $N_f=2$ flavours \cite{Drach:2020wux}.

In this work, we consider an $SU(2)$ gauge theory with $N_f = 2$ fundamental
Dirac fermions. The theory features an extended $SU(4)$ flavour
symmetry that spontaneously breaks to $Sp(4)$.  The theory is used to
build a pseudo-Nambu--Goldstone boson (PNGB) Composite Higgs model in Ref.~\cite{Cacciapaglia:2014uja}, and it was recently reviewed
in Ref.~\cite{Cacciapaglia:2020kgq}. In this model, the physical Higgs boson
is a mixture of PNGBs and of the flavour singlet state of the strong
sector. The model has been shown to pass experimental constraints~\cite{Cacciapaglia:2020kgq}, and the mixing between the scalar
resonance and the Higgs can relax the bounds on the model~\cite{BuarqueFranzosi:2018eaj}.

We present here the first calculation of the scattering amplitude of Goldstone bosons in the flavour singlet channel beyond
QCD. We have used two operators to constrain the scattering amplitude at two different kinematic configurations. The evaluation of disconnected
contributions increases significantly the computational cost with respect to other channels.

We also report on the comparison of our results to the chiral perturbation theory predictions (in isolation of the SM) of Ref.~\cite{Bijnens:2011fm}, which should match in the limit of light enough PNGBs.

\section{Lattice setup}
\label{sec:lattice}

We use the HiRep \cite{DelDebbio:2008zf} suite to simulate an $SU(2)$ gauge theory with $N_f=2$. For the fundamental fermions the action of choice is the Wilson action~\cite{Wilson:1974sk} with tree-level $O(a)$-improvement clover term~\cite{Sheikholeslami:1985ij}. For the gauge we use the tree-level Symanzik improved action~\cite{Luscher:1984xn}. Both the bare mass term, $ a \,m_0$, and the Wilson term explicitly break the $SU(4)$ flavour symmetry to an $Sp(4)$ subgroup.
All of our simulation are performed with periodic boundary conditions in all space-time directions, both in gauge and fermion\footnote{In $SU(2)$, periodic and antiperiodic boundary conditions differ only by a gauge transformation. } fields.

\begin{table}[h!]
\begin{tabular}{c|c|c|c|c|c|c}
Ensemble  & $L/a$  & $T/a$ & $\beta$ & $a \,m_0$ & $c_{sw}$  & \# configs  \\ \hline \hline
Heavy & 24 & 48 & 1.45 & $- 0.6050$ & 1.0 & 1980\\ \hline
Light  & 32 & 48 & 1.45 & $- 0.6077$ & 1.0 &  1160\\ 
\end{tabular}
\caption{ Simulation parameters in our ensembles. \label{tab:ensembles}}
\end{table}

The ensembles used for this work have been generated for $\beta=1.45$, and two different values of the bare fermion mass. We refer to these ensembles as ``light'' and ``heavy'' depending on the value of the pion mass. Here and in the following, we will make use of the naming convention inherited from QCD, that is, the pseudoscalar PNGB of this theory is referred to as pion. The spatial size of the ensembles has been tuned to obtain a value of $M_\pi L \simeq 5$. All the relevant simulation parameters are given in Table~\ref{tab:ensembles}.

 For each ensemble, we compute the PNGB mass, $M_\pi$, and the vector mass, $M_\rho$ from the Euclidean time dependence of appropriate correlation functions. We also extract the bare pseudoscalar decay constant $F_\pi^{\rm{bare}}$, which renormalises multiplicatively with the renormalization factor $Z_A$.  For more details about the calculation of these quantities, we refer the reader to Ref.~\cite{Arthur:2016dir}.  All our findings are summarised in \tab{tab:pionmass}.

In addition, the non-perturbative determination of $Z_A$ was carried out using the RI'-MOM scheme~\cite{Martinelli:1994ty}, using the same strategy as in the previous setup~\cite{Arthur:2016dir}. For detailed information about the $Z_A$ determination we refer to Ref.~\cite{Drach:2020wux}, where we estimated $Z_A=0.8022(3)$ for the same value of $\beta$ used in this work.

\begin{table}[h!]
\begin{tabular}{c|c|c|c|c}
Ensemble  &  $a M_\pi$ & $a M_\rho$  & $aF_\pi^\text{bare}$ & $M_\pi/F_\pi^\text{bare}$   \\ \hline \hline
Heavy & 0.2065(12) & 0.438(27) & 0.0395(9) & 5.24(11)  \\ \hline
Light &  0.1597(18) & 0.3864(30) &  0.0357(9) &  4.36(11) \\ 
\end{tabular}
\caption{ Pion mass, vector mass and decay constant for our two ensembles. \label{tab:pionmass}}
\end{table}

\section{Scattering in $\boldsymbol{SU(2)}$}
\label{sec:scattering}

In this section we will review and extend the necessary theoretical background
for this work. In particular, we will derive all the group classification
needed to evaluate the operators and the associated correlation
functions for the singlet channel, as well as the finite-volume
scattering formalism and the effective field theory (EFT) description
of the relevant scattering amplitude.

\subsection{Flavour singlet operators} 

We start by considering the flavour symmetries of the $SU(2)$ gauge
theory with $N_f=2$. It can be shown that the massless Lagrangian is
symmetric under an $SU(4)$ flavour transformation, while the mass term
can be shown to be $Sp(4)$ invariant. This means that there exist five
broken generators, which correspond to the pseudo-Nambu--Goldstone
fields. More specifically, it can be shown that they correspond to the
three pions and two dibaryons. In terms of the two fundamental fermion  fields $u$ and $d$, we can construct one-particle operators with the right quantum numbers as follows:
\begin{align}\label{eq:def_pi}
\Pi_{ud}(x) &= u^T(x) (-i \sigma_2) C \gamma_5 d(x), \nonumber \\
 \Pi_{\bar u \bar d}(x)  &=  \bar u(x) (-i \sigma_2) C \gamma_5 \bar d(x)^T,  \nonumber\\
  \pi^-(x) &=  \bar u(x) \gamma_5 d (x), \\
  \pi^+(x) &= - \bar d(x) \gamma_5 u(x),  \nonumber\\
\pi^0(x) &=\frac{1}{\sqrt{2}} \left[ \bar u(x) \gamma_5 u(x)  -  \bar d(x) \gamma_5 d(x)  \right], \nonumber
\end{align}
where $x\equiv ( {\bf x}, t)$. The real and antisymmetric matrix ($-i
\sigma_2$) acts in colour space, and $C$ represents the conjugation charge matrix, $C= i \gamma_0
\gamma_2$.
As we are interested only in the flavour structure of the operators, we will omit the space-time dependence of the fields in
the equations where possible.

In order to build a flavour singlet operator, we introduce:
 \begin{align}\label{eq:Q}
\begin{split}
Q = \begin{pmatrix} 
u_L \\
d_L\\
\widetilde{u}_L\\
\widetilde{d}_L\\
\end{pmatrix} &= \begin{pmatrix} 
u_L \\
d_L\\
(-i\sigma_2) C \bar{u}^T_R\\
(-i\sigma_2) C \bar{d}^T_R\\
\end{pmatrix},\\ \quad E &= \begin{pmatrix} 0 & \mathbbm{1}_2 \\ -\mathbbm{1}_2  & 0 \end{pmatrix} \,,\end{split}
\end{align}
where we are using the convention from Ref.\cite{Ryttov:2008xe}, summarised in
appendix \ref{app:conv}, together with the standard definition of
$q_{L,R} = P_{L,R}\, q$ and $\bar{q}_{L,R} = \bar{q}\, P_{R,L}$ where $P_L ={(1-\gamma_5)}/{2}$ and $P_R
= {(1+\gamma_5)}/{2}$.

With the above convention we can define  the multiplet   $\Pi^{i=1,\dots,5}$ and the singlet $\mathcal{O}_\sigma$  as  
\begin{align}\label{eq:GBs}
\begin{split}
\Pi^i &=\frac{1}{2} \left[ Q^T (-i\sigma_2) C \gamma_5 X^i E Q +
  \textrm{h.c}\right] \ , \\
 \qquad {\mathcal O_\sigma} &= \frac{1}{\sqrt{2}} \left[Q^T
   (-i\sigma_2)   C E  Q +   \textrm{h.c} \right]. 
\end{split}
\end{align}

Here  $X^{i=1,\dots,5}$ are the broken generators used to
parametrise the coset $SU(4)/Sp(4) $ defined in the appendix.

Considering the infinitesimal transformation
\begin{equation}
Q \longrightarrow\left(\mathbbm{1}_4 + i \alpha^a S^a \right) Q,
\end{equation}
where $\alpha^{i=1,\dots,10}$ are real parameters, and
$S^{a=1,\dots,10}$ are the generators of the Lie Algebra of $Sp(4)$. The generators obey
the Lie algebra defining relation:
\begin{equation}
E S^a + (S^{a})^T E = 0.
\end{equation}
It is straightforward to show that $\mathcal{O}_\sigma$ is a singlet of $Sp(4)$.   It
can also be shown by performing explicitly an infinitesimal
transformation that the multiplet ${\Pi}$ transforms as a 5-dimensional irreducible
representation of $Sp(4)$ and that any operator proportional to $ \tr{\Pi \otimes
  \Pi}$ is a singlet of $Sp(4)$. The reader interested in more details
is referred to Appendix \ref{app:flavour}.\\ The  operator 
\begin{align}
\mathcal{O}_{\pi\pi}=  -\frac{4}{\sqrt{5}}  \sum_{i=1}^5 \Pi^i \Pi^i 
\end{align}
is therefore a flavour singlet operator. Expressing the operator $\mathcal{O}_{\pi\pi}$ in terms
of the bilinear defined in Eq.~\ref{eq:def_pi}, we find:
\begin{align}
\begin{split}
\mathcal{O}_{\pi\pi} = \frac{1}{\sqrt{5}} \Big[   &
  +\pi^+ \pi^- +  \pi^- \pi^+ -\pi^0 \pi^0\\
&+ \Pi_{ud} \Pi_{\bar u \bar  d}    +
  \Pi_{\bar u \bar d} \Pi_{ud} \Big].
\end{split}
\end{align}
Similarly the operator $\mathcal{O}_{\sigma}$ can be expressed in
terms of the $u$ and $d$ fields as:
\begin{align}
\mathcal{O}_{\sigma} =  \frac{1}{\sqrt{2}} \left[ \bar u(x) u (x) + \bar d(x) d (x)\right].
\end{align}

In the following, we will use $\mathcal{O}_{\pi\pi}$ and
$\mathcal{O}_\sigma$ as the relevant operators to study the singlet
channel. We refer to them respectively as the two-pion and sigma operators.

\subsection{Contractions}

\begin{figure*}[ht]
\begin{picture}(0.7\textwidth,135)
\put(0,77){\includegraphics[width=0.7\textwidth,clip]{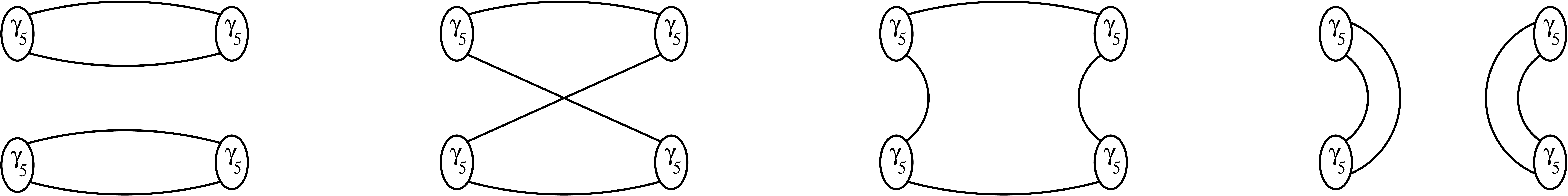} }
\put(0,10){\includegraphics[width=0.7\textwidth,clip]{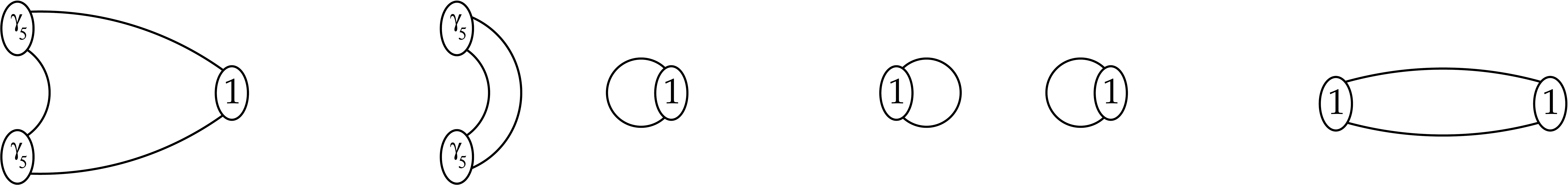} }
\put(17,68){$D(t)$}
\put(117, 68){$X(t)$}
\put(219, 68){$R(t)$}
\put(320, 68){$V(t)$}
\put(17,0){$T(t)$}
\put(117,0){$W(t)$}
\put(219,0){$\Sigma(t)$}
\put(320,0){$B(t)$}
\end{picture} 
\caption{Representation of the different contractions needed for this work. The blobs indicate a fermion bilinear, with gamma matrix $\gamma_5$ or identity. The physical correlation functions are constructed from linear combinations thereof as given in Eq.~\ref{eq:contractions}.  }
\label{fig:contractions} 
\end{figure*}
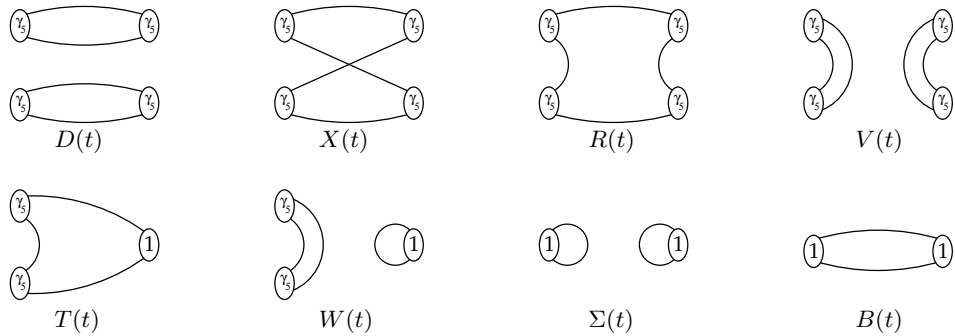

In the rest of the paper we will use the zero momentum projection of the operators
defined in Eq.~\ref{eq:def_pi} for the evaluation of the correlators. Explicitly, this is given by
\begin{align}
\Pi_{ud}(t) &= \sum_{\bf x} \Pi_{ud}({\bf x},t),
\end{align}
and analogously for the other one-particle operators.

The energy of the flavour singlet state can be computed from the
exponential decay in time of the appropriate correlation functions of
the two-pion and sigma operators described in the previous section.

The singlet two-pion operator, with each one-particle operator projected at zero momentum is
  \begin{align}
\begin{split} 
\mathcal O_{\pi\pi}(t) &= \frac{1}{\sqrt{5}} \bigg[  \pi^+(t) \pi^-(t) +   \pi^-(t) \pi^+(t)   \\ 
& - \pi^0(t) \pi^0(t) \\& + \Pi_{ud} (t) \Pi_{\bar u \bar d} (t) + \Pi_{\bar u\bar d} (t) \Pi_{ u  d} (t)
 \bigg],  \label{eq:Oppipi}
\end{split}
  \end{align}
 where we have included the Euclidean time explicitly. Analogously,
 the zero momentum projected sigma operator can be rewritten as
\begin{equation}
\mathcal O_{\sigma}(t) = \frac{1}{\sqrt{2}}  \sum_{\bf x} \left(\bar u({\bf x},t) u({\bf x},t) + \bar d( {\bf x},t) d({\bf x},t)\right). \label{eq:Opsigma}
\end{equation}
Using the two operators in Eqs.~\ref{eq:Oppipi} and \ref{eq:Opsigma}, we can build a symmetric two-by-two matrix of correlation functions as follows:
\begin{equation}
C_{X \to Y} (\delta t)= \frac{1}{T} \sum_{t} \langle O_X(t+\delta t) O_Y(t)^\dagger\rangle. \label{eq:matrixC}
\end{equation}
By solving the associated generalised eigenvalue problem
(GEVP)~\cite{Luscher:1990ck}, we are able to obtain the energy of the
two lowest states in the spectrum, by  measuring the exponential decay of the two eigenvalues.

The three different correlation functions that enter in Eq.~\ref{eq:matrixC} can be built from eight different Wick contractions:
\begin{align}
\begin{split}
C_{\sigma \to  \sigma}(t) &= -B(t) + 2 \Sigma(t),  \\
C_{\pi\pi \to  \pi\pi}(t) &= 2 D(t) + 3 X(t) - 10 R(t) + 5 V(t), \\
C_{\pi\pi \to  \sigma}(t) &= \sqrt{10} \left( T(t)- W(t) \right). \label{eq:contractions}
\end{split}
\end{align}
These are defined in Fig.~\ref{fig:contractions}, along with their
naming conventions. Three of the contractions include disconnected
diagrams: $V$, $W$ and $\Sigma$, and, as will be seen later, they dominate the statistical uncertainty.

\subsection{Extraction of scattering amplitudes}
\label{sec:phaseshift}

The Lüscher method~\cite{Luscher:1986pf,Luscher:1990ux,Luscher:1991cf}
provides a way to obtain two-particle scattering amplitudes from
lattice simulations. The so-called quantization condition connects the
finite-volume energy levels to the phase shift. It is a
well-established technique~\cite{Rummukainen:1995vs, Kim:2005gf,
  He:2005ey, Bernard:2010fp, Briceno:2012yi, Briceno:2014oea,
  Romero-Lopez:2018zyy,Luu:2011ep,Gockeler:2012yj}, which has been
applied to many systems---see Ref.~\cite{Briceno:2017max} for a
review.
In the context of QCD the singlet channel has often been studied, see
for example Refs.~\cite{Guo:2018zss,Fu:2017apw,Mai:2019pqr,Briceno:2016mjc,Briceno:2017qmb,Liu:2016cba}.

In the case of two identical scalars with only $s$-wave interactions, the quantization condition reads~\cite{Luscher:1986pf}:
\begin{equation}
k \cot \delta_0 \left( k \right) = \frac{2}{\sqrt{\pi} L} \mathcal Z_{00}(\eta^2), \quad \eta = \frac{L k}{2 \pi}, \label{eq:QC}
\end{equation}
where the energy levels are in the $A_1^+$ irreducible representation of the octahedral group, and $k$ is the relative momentum in the center-of-mass (CM) frame.
Furthermore, $\mathcal Z_{00}$ is the standard L\"uscher zeta
function. Note that in this form, the quantization condition is a
one-to-one mapping between an energy level and a point in the phase
shift curve.

It is convenient, for our discussion later, to highlight how bound
states manifest themselves in the phase shift both at finite and infinite volume.

In infinite volume, they correspond to poles in the scattering amplitude. The pole's position is given by 
\begin{equation}
k \cot \delta_0\left( k \right)= - \sqrt{-k^2 },
\end{equation}
which we denote as bound-state condition. The fact that the residue of the pole has a positive sign, implies the following condition~\cite{Iritani:2017rlk}:
\begin{equation}
\frac{d}{dk^2} \left[ \ k \cot \delta_0 \left( k \right) - \left(- \sqrt{-k^2} \right) \ \right] < 0.
\end{equation}
This means that $k \cot \delta_0$ must cross the bound-state condition from below with decreasing $k^2$.  

By contrast, the finite-volume solutions to the quantization condition
never intersect the bound-state condition. They are however
exponentially close~\cite{Luscher:1985dn}, with an exponent related to
the binding momentum~\cite{Konig:2017krd}.

\subsection{EFT prediction}

At sufficiently low energies and close to the chiral limit, Chiral
Perturbation Theory (ChPT) should provide a satisfactory description
of the interactions of Goldstone bosons in QCD-like theories. However,
the precise predictions depend upon the symmetry breaking pattern. As
explained before, in our case an $SU(4)$ flavour symmetry is
spontaneously broken down to $Sp(4)$. This was worked out in
Refs.~\cite{Bijnens:2009qm,Bijnens:2011fm}, and is referred to as the
pseudo-real case.

In the present work the quantity of interest is the two-pion
scattering amplitude in the singlet channel---analogous to that of the
``$\sigma$'' resonance in QCD. In this exploratory study, the leading-order (LO) ChPT result will suffice. This reads

\begin{equation}
\mathcal T_I = \frac{M_\pi^2}{F_\pi^2}  \left(-\frac{3}{2} + 2 \frac{s}{M_\pi^2} \right),
\end{equation}
where we are using the convention  $f_\pi=\sqrt{2} F_\pi$ for the
normalization of the decay constant, and $\sqrt{s}$ is the CM
energy. From the scattering amplitude, the momentum dependence of the phase-shift can be easily derived:
\begin{equation}
\text{Re }\frac{1}{\mathcal T_I }= \frac{k \cot \delta^I_0}{16 \pi \sqrt{s}}.
\end{equation}
The LO result is
\begin{equation}
\frac{k}{M_\pi} \cot \delta^{I}_0 =  \frac{M_\pi \sqrt{s}}{13 M_\pi^2 + 16 k^2} \left( \frac{32 \pi  F_\pi^2}{M_\pi^2}  \right). \label{eq:kcot}
\end{equation}
Furthermore, the scattering length is defined as
\begin{equation}
\lim_{k \to 0} \frac{k}{M_\pi} \cot \delta^I_0  = -\frac{1}{M_\pi a^I_0},
\end{equation}
and its result reads
\begin{equation}
M_\pi a_0^I = - \frac{13}{64\pi}  \frac{M_\pi^2}{F_\pi^2}.
\end{equation}
An interesting remark is that the leading-order amplitude has a zero below threshold (Adler zero), which translates to a pole in $k \cot \delta_0^I$. This is located at $(k/M_\pi)^2 = -13/16$, and may limit the converge of a polynomial expansion of $k \cot \delta_0^I$ in $k^2$---the so-called threshold expansion. Such behaviour has been observed, e.g., in the isospin-2 $\pi\pi$ system in QCD~\cite{Blanton:2019vdk}.

\section{Results}
\label{sec:results}

\subsection{Correlation functions}
\label{sec:corr}

We construct the correlation functions as indicated in Eq.~\ref{eq:contractions}. In order to evaluate all the contractions depicted in Fig.~\ref{fig:contractions}, we use various types of stochastic sources. First, for the $D, X, R, V$ and $B$ contractions we use time-diluted stochastic sources. By placing a source in each of the timeslices, we can obtain a single stochastic estimator for each of these contractions. In this case we use $10$ stochastic estimators, which require $T \times 10$ inversions of sources. By contrast, we use $40$ volume sources for the $\Sigma$ contraction, while for $W$ we combine the building blocks of $V$ and $\Sigma$. Finally, $T$ is computed by employing $40$ time-diluted sources that have an additional sequential inversion.

The contractions $\Sigma$ and $W$ are responsible for the largest contribution to the statistical uncertainty. This is because they contain the trace of a single propagator multiplied by the identity in spinor space, and so, they are dominated by the gauge noise. Because of this, we choose to measure them more often than the other building blocks. In fact, we measure the trace of the single propagator in steps of one unit of Monte Carlo time.

We perform the analysis of uncertainties using jackknife samples. In order to account for autocorrelations, we use the binning procedure. For this, we average correlation functions within a bin length of $10$ units of Monte Carlo time. We have checked that larger bin sizes, 20 and 30, do not lead to any substantial change in the estimation of uncertainties.

As the operators have vacuum quantum numbers, there is an overall constant in all our correlation functions. Because of this, we will work with the shifted correlator:
\begin{equation}
\widetilde C(t) = \frac{1}{2} \big[ C(t-1)-C(t+1) \big].
\end{equation}
This is a discrete version of the derivative in Euclidean time that keeps the same exponential decay, but cancels the undesired constant. 

\begin{figure}[h!]
   \centering
   \subfigure[ Ensemble with heavier pion mass. ]%
             {\includegraphics[width=0.475\textwidth,clip]{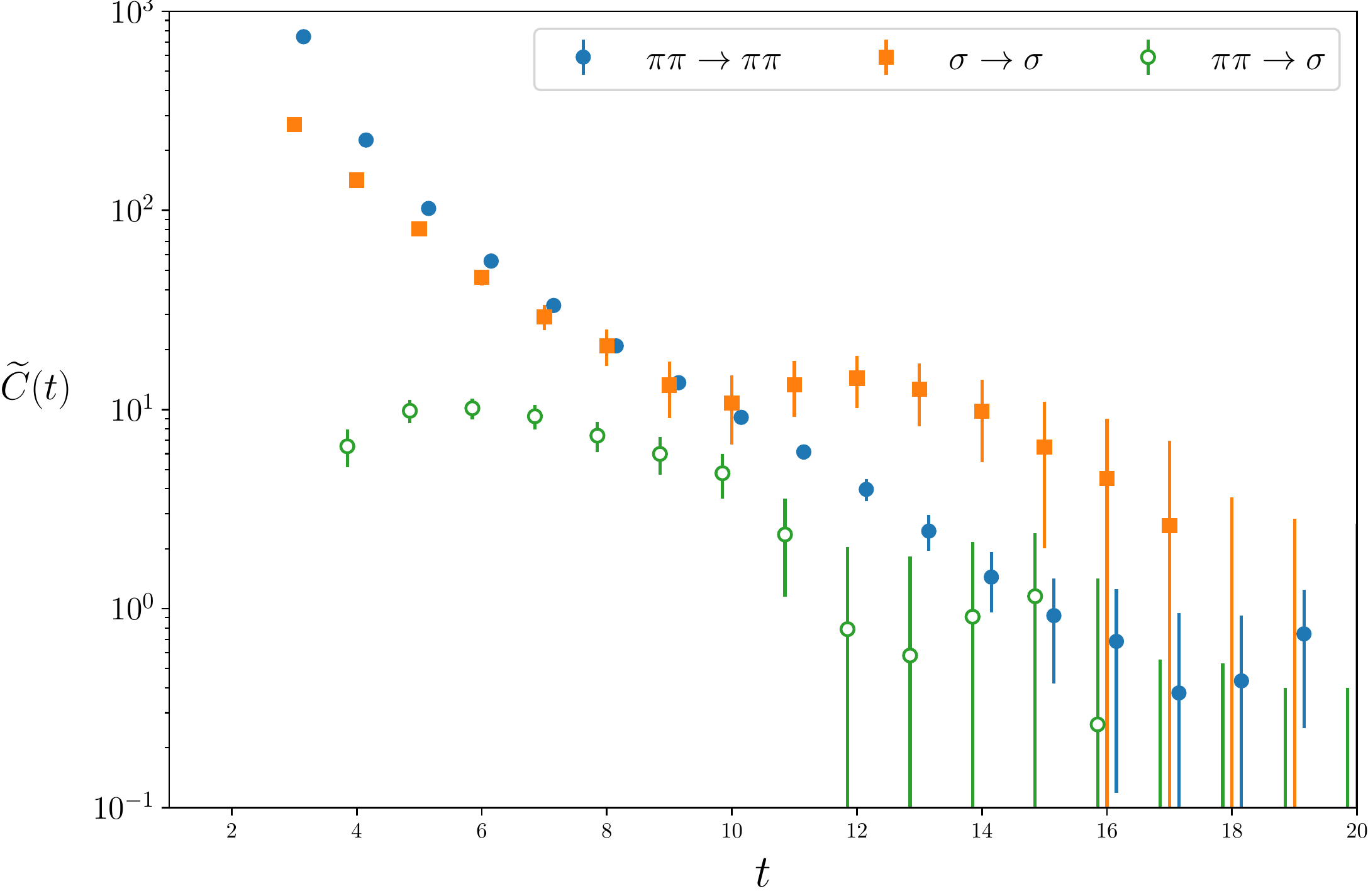} }\hfill
   \subfigure[  Ensemble with lighter pion mass. ]%
             {\includegraphics[width=0.475\textwidth,clip]{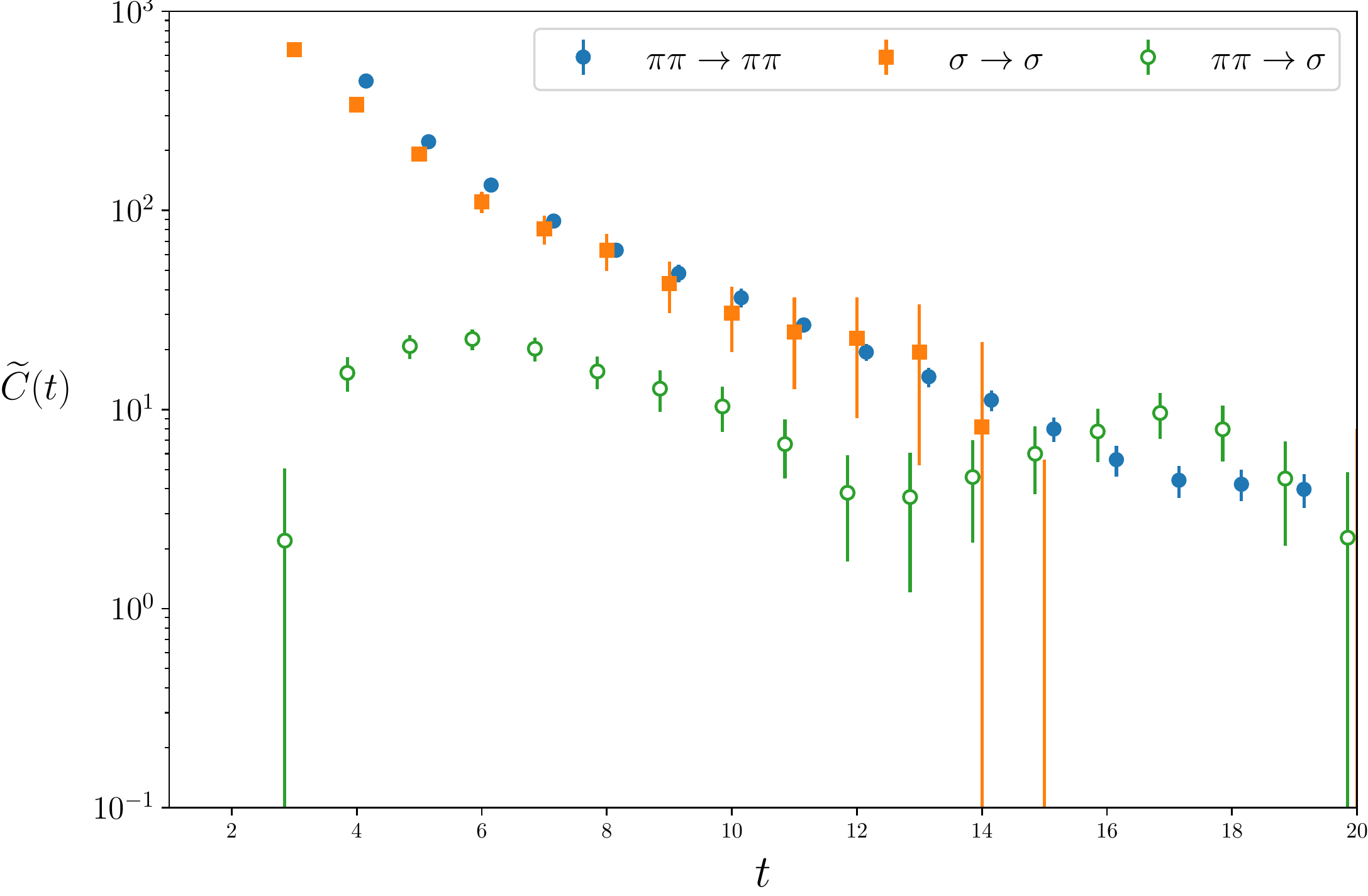} }
   \caption{Correlation functions built with two different operators with singlet quantum numbers. For visualization purposes we include an arbitrary normalization.}
\label{fig:correlators} 
\end{figure}

The results for the two ensembles are shown in Fig.~\ref{fig:correlators}.  As can be seen, the statistical noise is dominated by the ones including the $\mathcal{O}_\sigma$ operator, which contain the $W$ and $\Sigma$ contractions in Fig.~\ref{fig:contractions}. It is also clear that one cannot trust the correlator in the region dominated by the statistical noise.

\subsection{Spectrum determination}

We now turn to the determination of the spectrum. For this, we build a two-by-two matrix, as presented in Eq.~\ref{eq:matrixC}. The GEVP is defined by means of the shifted correlator as
\begin{equation}
\widetilde C(t) v_n (t,t_0) = \lambda_n (t,t_0) \widetilde C(t_0) v_n (t,t_0),
\end{equation}
where $t_0$ is a reference timeslice. Note that $\lambda_n$ are the eigenvalues of $\widetilde C^{-1}(t_0)  \widetilde C(t)$. In our case, we choose $t_0=4$ as for both ensembles it is the first stable point.

There are various ways of solving the eigenvalue equation. One can fix the diagonalisation point, or diagonalise separately in each timeslice. We opt for the latter, but we have seen that it does not lead to any substantial change compared the the other method. Regarding the estimation of uncertainties, we choose to diagonalise in each jackknife sample separately. We have also checked that fixing the diagonalisation in all samples barely alters the outcome.

 \begin{figure}[h!]
   \centering
   \subfigure[ Ensemble with heavier pion mass. ]%
             {\includegraphics[width=0.475\textwidth,clip]{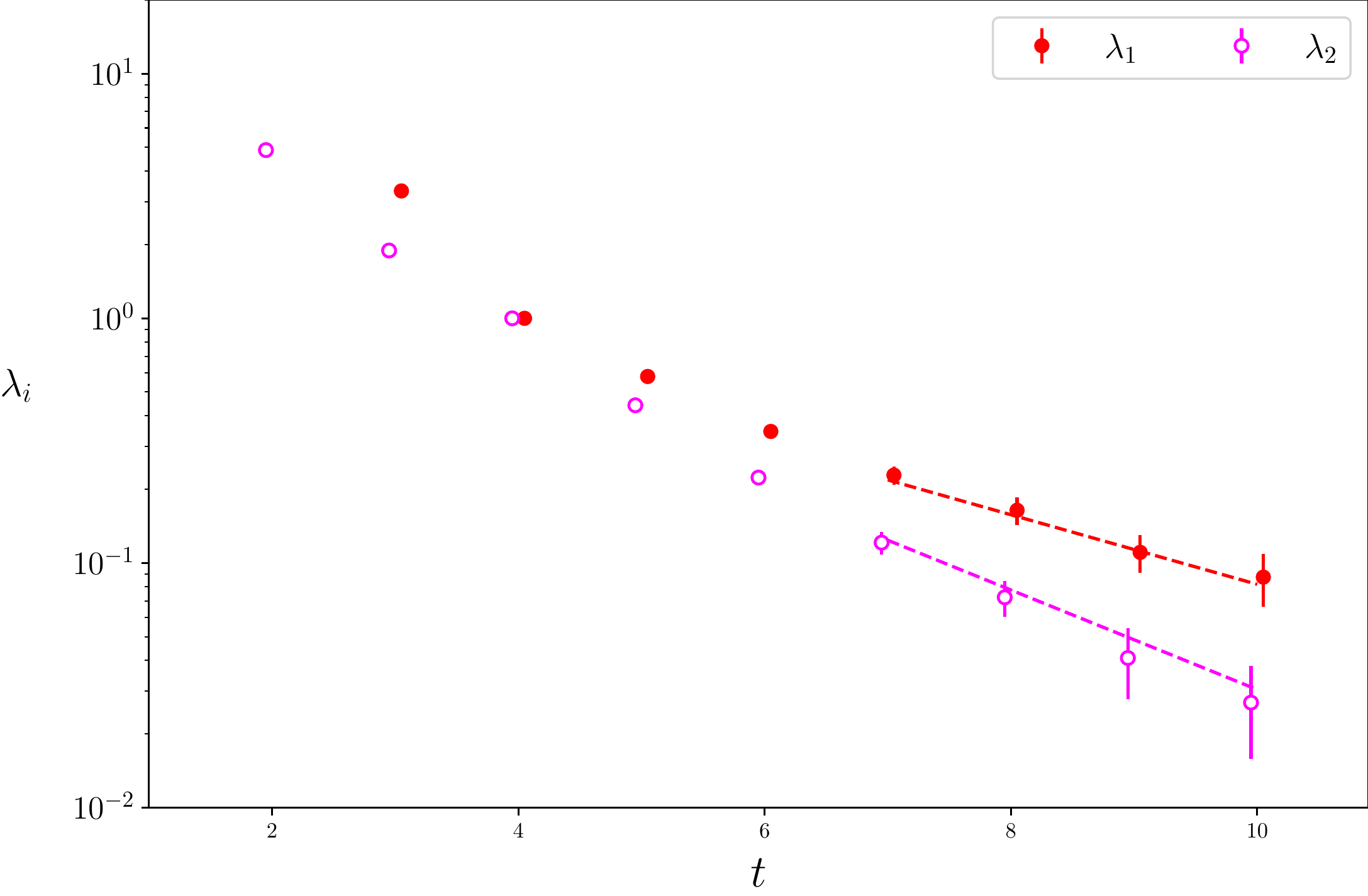} }\hfill
   \subfigure[  Ensemble with lighter pion mass. ]%
             {\includegraphics[width=0.475\textwidth,clip]{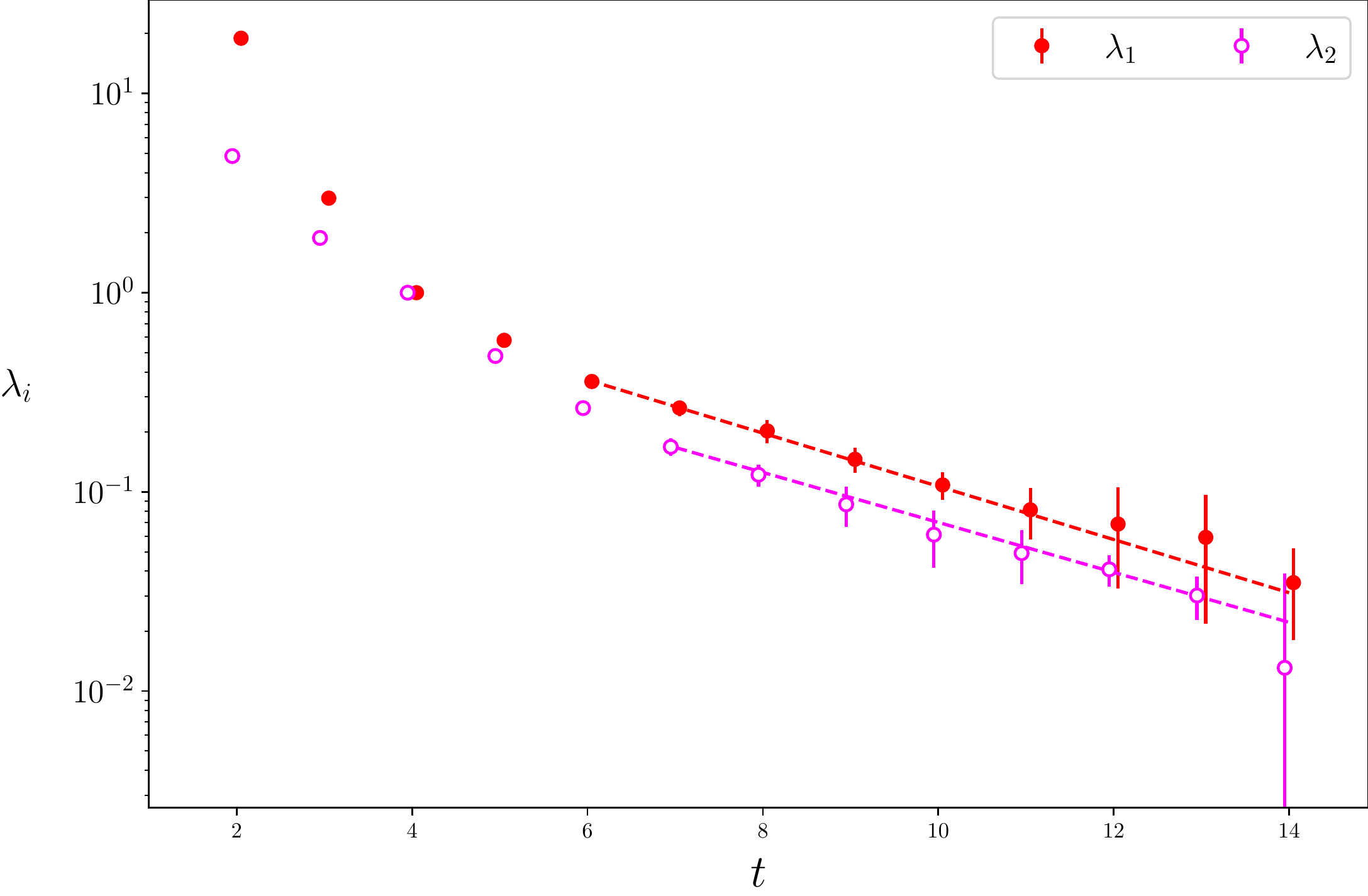} }
   \caption{Lowest two eigenvalues for the two ensembles of this work. The dashed line indicates the fit range.}
\label{fig:eigenvalues} 
\end{figure}

The dependence of the eigenvalues with Euclidean time is expected to be a sum of exponentials. Solving the GEVP allows one to isolate the low-lying energy states. In the limit of sufficiently large Euclidean time, each eigenvalue decays as a single exponential:
\begin{equation}
\lambda_i(t) \longrightarrow A_i e^{-E_i t },
\end{equation}
which holds up to effects that are exponentially suppressed with the time extent of the lattice---thermal effects. The corresponding exponents, $E_i$, are associated to one energy level of the studied channel.

 The dependence of the eigenvalues with Euclidean time is shown in Fig.~\ref{fig:eigenvalues}. The dashed lines depict the best fit in the chosen fit interval. Note that we do not include in the fits the region in which the $\widetilde C_{\sigma\sigma}$ correlator is dominated by noise.

\begin{figure}[h!]
   \centering
             {\includegraphics[width=0.5\textwidth,clip]{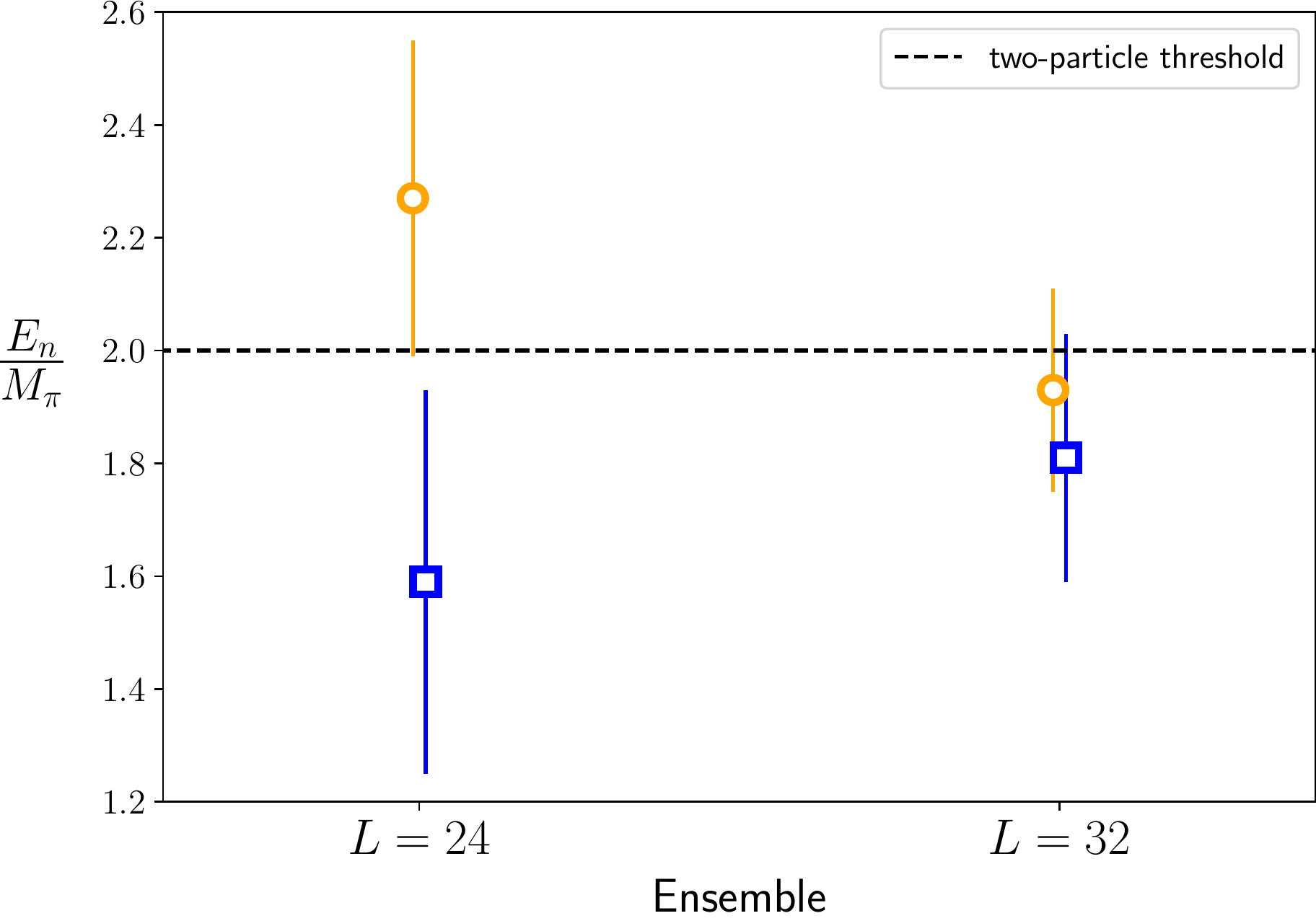} }\hfill
   \caption{Energy levels obtained from our simulations. The blue square represents the lowest state, and the orange circle the first excited state.}
\label{fig:spectrum} 
\end{figure}

A summary of the extracted energy levels---in units of the pion mass---is given in Tab.~\ref{tab:spectrum} and in Fig.~\ref{fig:spectrum}. As can be seen, the central value of the lowest energy is well below threshold, and the second level is around the two-particle threshold in both cases. The physical interpretation of these states can only be discussed after inspecting the scattering amplitude. In particular, to answer whether the lowest state corresponds to a bound state, or an attractive scattering state. This will be addressed in the next subsection.

\begin{table}[h!]
\begin{tabular}{c|c|c|c|c|c|c}
Ensemble   & $E_1/M_\pi$  &$\chi^2_{\rm red}$& $[t_i , t_f]$ &$E_2/M_\pi$  & $\chi^2_{\rm red}$ & $[t_i , t_f]$  \\ \hline \hline
Heavy & 1.59(34) &1.47  &  [7,10]&2.27(28)&0.25 & [7,10] \\ \hline
Light  & 1.81(22)  &  0.18 & [7,14]  & 1.93(18)  & 0.30& [6,14] \\ 
\end{tabular}
\caption{ Two-particle energy levels in the singlet channel extracted from the fits in Fig.~\ref{fig:eigenvalues}. We show the $\chi^2$ per degree of freedom, $\chi^2_{\rm red} = \chi^2/\text{dof}$, and the fit range for each level. \label{tab:spectrum}}
\end{table}

\subsection{Results for the scattering amplitude}
\label{sec:resphaseshift}

We are now in the position to explore the scattering amplitude in the singlet channel using the Lüscher method. For this, we insert the energy levels of Tab.~\ref{tab:spectrum} into the two-particle quantization condition in Eq.~\ref{eq:QC}. 

\begin{figure*}[tpb]
   \centering
   \subfigure[ \, Ensemble with heavier pion mass.]%
             {\includegraphics[width=0.475\textwidth,clip]{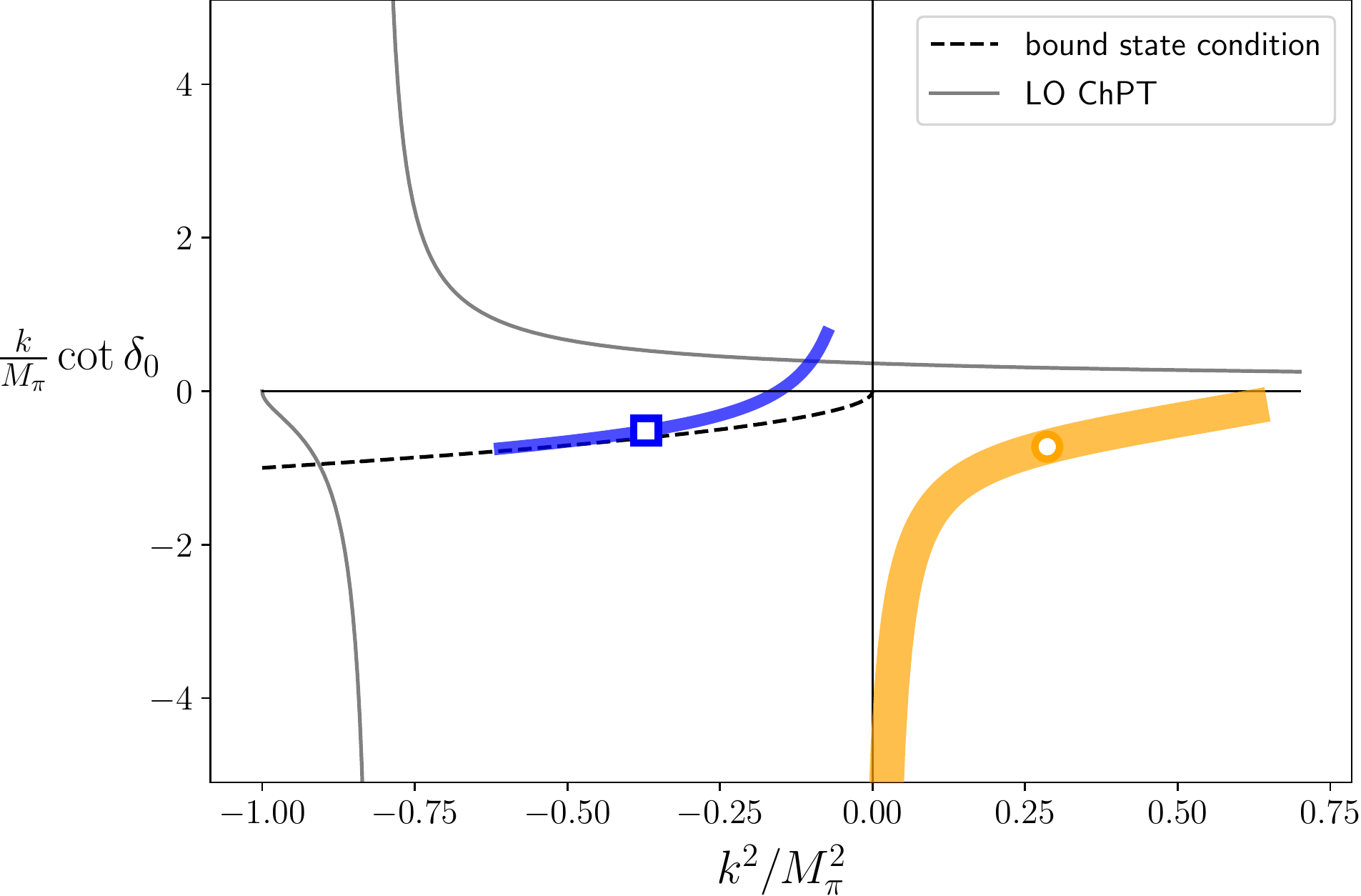} }\hfill
   \subfigure[ \, Ensemble with lighter  pion mass.]%
             {\includegraphics[width=0.475\textwidth,clip]{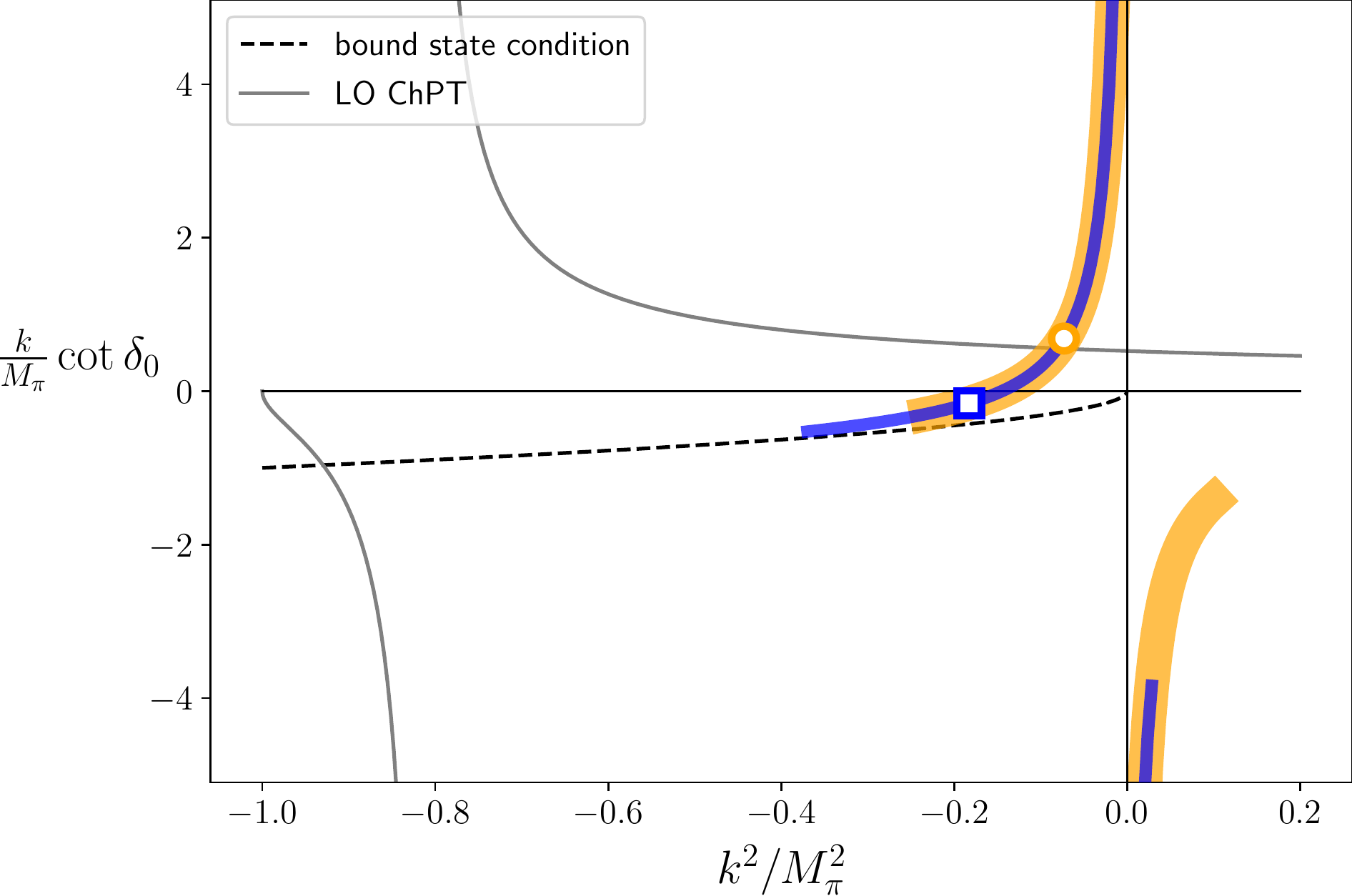} }
   \caption{$s$-wave phase shift in the form $(k/M_\pi) \cot \delta_0$ for the two ensembles of this work. The empty marker is the central value, as the shaded area represents the $1\sigma$ resulting from the quantization condition. The blue squares correspond to the lowest state, and the orange circles to the first excited state (same convention as in Fig.~\ref{fig:spectrum}) . Note that the width of the shaded area is arbitrary, and has been chosen for illustrative purposes. We also include the leading-order chiral prediction, as well as the bound-state condition. }
\label{fig:kcot} 
\end{figure*}

The corresponding points in the phase shift are shown in Fig.~\ref{fig:kcot} for both ensembles, where we also include the $1\sigma$ region for visualization.  As can be seen in this figure, we find in both cases a point below threshold whose central value is close to the bound state condition. Even if the uncertainty is large, the most likely interpretation is that there is indeed a bound state in this channel for the explored pseudoscalar mass. The second point in the curve is around threshold, and thus could be used to constrain the scattering length of the channel. Unfortunately, the uncertainty is too large and the result is inconclusive.

We can also comment on the comparison of our results and the leading-order prediction from ChPT. This is  depicted as solid grey line in Fig.~\ref{fig:kcot}. The LO ChPT prediction shows no sign of a bound state in the region where we seem to find one. It does however predict one bound state well below threshold, which is an artefact caused by the Adler zero~\cite{GomezNicola:2007qj}. Moreover, the leading chiral prediction is also not able to accommodate the observed points around threshold. Thus, it seems that the value of the pseudoscalar masses of our simulations are outside of the window for which leading-order ChPT is a good description. 

\section{Conclusion and Outlook}
\label{sec:conclu}

This work represents the first study of the singlet channel in four-dimensional gauge theories beyond QCD. Specifically, we have considered an $SU(2)$ gauge theory with two fundamental fermions that serves as a minimal template for a Composite Higgs model.

In this theory, the symmetry breaking pattern differs from that of QCD---the $SU(4)$ flavour symmetry breaks down to $Sp(4)$. Therefore, we have derived the group-theoretical setup required to analyse this scattering channel. It can also be noted that our analysis holds for generic $Sp(2N)$ gauge theories with two fundamental fermions, as the same symmetry breaking pattern is realised.

We have used two ensembles with different pion masses. Using two different operators to solve the GEVP, we have computed the lowest two energy levels. These are fed into the Lüscher quantization condition, and we have been able to put non-perturbative constraints on the singlet scattering amplitude. Interestingly, we find that leading-order chiral perturbation theory does not seem to describe the amplitude correctly, and fails in predicting a bound state around the region where we observe it.

Our results strongly suggest that in the explored region of fermion masses the sigma is most likely a stable particle, that is, a two-pion bound state. In our two ensembles, we find that $M_\sigma/M_\pi \sim 1.5-1.8$. We however expect this feature to depend strongly upon the pion mass. Therefore, more work is required to investigate discretisation effects and to reach the phenomenologically appealing region, that is, where the sigma becomes unstable. We expect to pursue this direction in a subsequent work.

\vspace{0.5cm}

\begin{acknowledgments}

We thank Laurence Bowes for comments on the manuscript.

The work of FRL has also received funding from the European Union Horizon 2020 research and innovation program under the Marie Sk\l{}odowska-Curie grant agreement No. 713673 and "La Caixa" Foundation (ID 100010434, LCF/BQ/IN17/11620044). FRL also acknowledges support from the Generalitat Valenciana grant PROMETEO/2019/083, the European project H2020-MSCA-ITN-2019//860881-HIDDeN, and the national project FPA2017-85985-P. FRL has also received financial support from Generalitat Valenciana through the plan GenT program (CIDEGENT/2019/040). 
The work has been performed under the Project HPC-EUROPA3
(INFRAIA-2016-1-730897), with the support of the EC Research
Innovation Action under the H2020 Programme.

AR and PF are supported by the STFC Consolidated Grant ST/P000479/1.
VD is supported by the STFC Consolidated Grant ST/T00097X/1.

This work was performed using the Cambridge Service for Data Driven
Discovery (CSD3), part of which is operated by the University of
Cambridge Research Computing on behalf of the STFC DiRAC HPC Facility
(www.dirac.ac.uk). The DiRAC component of CSD3 was funded by BEIS
capital funding via STFC capital grants ST/P002307/1 and ST/R002452/1
and STFC operations grant ST/R00689X/1. This work was also performed using the DiRAC Data Intensive service at
Leicester, operated by the University of Leicester IT Services, which
forms part of the STFC DiRAC HPC Facility (www.dirac.ac.uk). The
equipment was funded by BEIS capital funding via STFC capital grants
ST/K000373/1 and ST/R002363/1 and STFC DiRAC Operations grant
ST/R001014/1. DiRAC is part of the National e-Infrastructure.

This work used the ARCHER UK National Supercomputing Service (http://www.archer.ac.uk).

We thank the University of Plymouth for providing computing time on
the local HPC cluster.

\end{acknowledgments}

\newpage

\begin{appendix}
\section{Lie algebra of $\boldsymbol{SU(4)}$}\label{app:conv}
Following the convention used in Ref.~\cite{Ryttov:2008xe}, we define
\begin{align}
\begin{split}
B_1 &=  \sigma_4, B_2 =  i \sigma_4, B_3 =  \sigma_3, 
 B_4 =  i \sigma_3, \\ B_5 &=  \sigma_1, B_6 =  i \sigma_1, 
 D_4 = \sigma_2, D_5 = i\sigma_2\, ,
\end{split}
\end{align}
where $\sigma_4$ is the identity matrix, and $\sigma_{i=1,\dots,3} $ are
the Pauli matrices.  The ten generators of $Sp(4)$ are denoted  $S^{a=1,\dots,10}$, together
with the five broken generators  $X^{i=1,\dots,5}$ they are a basis of the
Lie Algebra of $SU(4)$. They are defined as follows:
\begin{align}
\begin{split}
S^{a} &= \frac{1}{2\sqrt{2}} \begin{pmatrix} \sigma_a &0 \\
0 & -\sigma^T_a\end{pmatrix},  \quad a = 1,\dots, 4 \\
 S^a &= \frac{1}{2\sqrt{2}}\begin{pmatrix} 0 & B_{a-4} \\
 B_{a-4}^\dagger& 0
\end{pmatrix}, \quad a = 5,\dots, 10   \\
  X^i  &= \frac{1}{2\sqrt{2}} \begin{pmatrix} \sigma_i &0 \\
0 & \sigma^T_i
\end{pmatrix},  \quad i = 1,\dots, 3 \\
X^i &= \frac{1}{2\sqrt{2}} \begin{pmatrix} 0 &D_i \\
 D_i^\dagger& 0
\end{pmatrix}, \quad i = 4, 5~. 
\end{split}
\end{align}
The generators $S^a$ satisfy the relation $(S^a)^T E + E S^a = 0 $. The generators are normalised so that:
\begin{align}
\begin{split}
\tr{ S^a S^b } = \frac{1}{2} \delta^{ab}, \quad &\tr{ X^i X^j } = \frac{1}{2} \delta^{ij}, \\ \tr{ S^a X^i }& = 0 \,.
\end{split}
\end{align}
The structure constants of the algebra of $Sp(4)$ are defined as $f_{abc} = 2 \tr{ S^a [S^b,S^c]}$.
 
\section{Transformation under the flavour symmetry group $\boldsymbol{Sp(4)}$}
\label{app:flavour}
Using the following relations:
\begin{align}
\begin{split}
\{C,\gamma_5 \} = 0, \quad C^T = -C, \\ \quad C^2 = 1, \quad (-i \sigma_2)^2 = -1,
\end{split}
\end{align}
and the definitions of the Goldstone
bosons interpolating fields in Eq.~\ref{eq:def_pi} we find that:
\begin{align}
\Pi &= \frac{1}{2} \left[Q^T (-i\sigma_2) C \gamma_5 X^i E Q +
  \textrm{h.c} \right]^{i=1,\dots,5} \nonumber \\
 &= \frac{1}{2 \sqrt{2}}\begin{pmatrix}
{\pi^- -\pi^+}  \\
i ({\pi^- +\pi^+}) \\
\sqrt{2} {\pi^0}  \\
i \left(\Pi_{\bar u \bar d} +\Pi_{ud} \right) \\
 {\Pi_{\bar u \bar d} -\Pi_{ud}} 
\end{pmatrix}.
\end{align}

Performing an infinitesimal transformation $Q \longrightarrow Q + i
\alpha^a S^a$ where $\alpha^a$ are real infinitesimal parameters, we find that 
\begin{align}
\begin{split}
\Pi &\longrightarrow \Pi + M \Pi, \quad \text{with} \\
M &=  \sqrt{2}\begin{pmatrix}
0 & - \alpha^3  & - \alpha^2 & -\alpha^7 & \alpha^8 \\
 \alpha^3  & 0 & - \alpha^1 &  -\alpha^6 & -\alpha^5 \\
 \alpha^2 &  \alpha^1 & 0 &  \alpha^9 &  -\alpha^{10} \\
 \alpha^7 &  \alpha^6 &  -\alpha^9 &0 &  \alpha^4 \\
- \alpha^8&  \alpha^5 &  \alpha^{10} &  -\alpha^4 & 0 
\end{pmatrix}.
\end{split}
\end{align}
Here, $M$ is an antisymmetric matrix that can be decomposed onto the algebra
of $SO(5)$, therefore showing that $\Pi$ belongs to a 5-dimensional
irreducible representation  of $Sp(4)$. The transformation of $\tr{ \Pi \otimes \Pi } $ therefore reads:
\begin{align}
\begin{split}
\tr{ \Pi \otimes \Pi } & \longrightarrow \tr{ \Pi \otimes \Pi } + \tr{
  M \Pi \otimes \Pi + \Pi \otimes M \Pi}   \\
&=  \tr{ \Pi \otimes \Pi } +  \Pi^T M \Pi  = \tr{ \Pi \otimes \Pi } ,
\end{split}
\end{align}where we have used that $M$ is antisymmetric in the last equality.
\end{appendix}

\vfill

  
\bibliography{biblio.bib}

\end{document}